\input harvmac.tex
\overfullrule=0pt

\def\simge{\mathrel{%
   \rlap{\raise 0.511ex \hbox{$>$}}{\lower 0.511ex \hbox{$\sim$}}}}
\def\simle{\mathrel{
   \rlap{\raise 0.511ex \hbox{$<$}}{\lower 0.511ex \hbox{$\sim$}}}}
 
\def\slashchar#1{\setbox0=\hbox{$#1$}           
   \dimen0=\wd0                                 
   \setbox1=\hbox{/} \dimen1=\wd1               
   \ifdim\dimen0>\dimen1                        
      \rlap{\hbox to \dimen0{\hfil/\hfil}}      
      #1                                        
   \else                                        
      \rlap{\hbox to \dimen1{\hfil$#1$\hfil}}   
      /                                         
   \fi}                                         %

\def\CL{{\cal L}}

\def\CO{{\cal O}}
\def\ts{\thinspace}
\def\ra{\rightarrow}

\def\Lra{\Longrightarrow}

\def\leftra{\leftrightarrow}

\def\ol{\bar}

\def\gev{{\rm GeV}}
\def\tev{{\rm TeV}}

\def\ecm{\sqrt{s}}

\def\half{\textstyle{ { 1\over { 2 } }}}
\def\third{\textstyle{ { 1\over { 3 } }}}
\def\fourth{\textstyle{ { 1\over { 4 } }}}
\def\twothirds{\textstyle{ { 2\over { 3 } }}}
\def\sixth{\textstyle{ { 1\over { 6 } }}}
\def\twelveth{\textstyle{ { 1\over {12} }}}

\def\Getc{G_{ETC}}

\def\suc{SU(3)_C}
\def\Ntc{N}
\def\sutc{SU(N)}
\def\uone{U(1)_1}
\def\utwo{U(1)_2}
\def\uy{U(1)_Y}
\def\suone{SU(3)_1}
\def\sutwo{SU(3)_2}
\def\suthree{SU(3)_3}

\def\condaa{\langle \bar T^1_L T^1_R\rangle}
\def\condbb{\langle \bar T^2_L T^2_R\rangle}
\def\condab{\langle \bar T^1_L T^2_R\rangle}

\def\condij{\langle \bar T^i_L T^j_R\rangle}

\def\condtbt{\langle \bar t t\rangle}

\def\tpi{\pi_T}
\def\trho{\rho_T}
\def\toppi{\pi_t}

\def\myfoot#1#2{{\baselineskip=14.4pt plus 0.3pt\footnote{#1}{#2}}}

\Title{\vbox{\baselineskip12pt\hbox{BUHEP--96--2}
\hbox{hep-ph/9602221}}}
\centerline{\titlefont{Symmetry Breaking and Generational Mixing}}
\vskip0.10truein
\centerline{\titlefont{in Topcolor--Assisted Technicolor}}

\bigskip
\centerline{Kenneth Lane\myfoot{$^{\dag }$}{lane@buphyc.bu.edu}}
\smallskip\centerline{Department of Physics, Boston University}
\centerline{590 Commonwealth Avenue, Boston, MA 02215}
\vskip .3in

\centerline{\bf Abstract}

Topcolor--assisted technicolor provides a dynamical explanation for
electroweak and flavor symmetry breaking and for the large mass of the top
quark without unnatural fine tuning. A major challenge is to generate the
observed mixing between heavy and light generations while breaking the
strong topcolor interactions near~$1\,\tev$. I argue that these phenomena,
as well as electroweak symmetry breaking, are intimately connected and I
present a scenario for them based on nontrivial patterns of technifermion
condensation. I also exhibit a class of models realizing this scenario.
This picture leads to a rich phenomenology, especially in hadron and lepton
collider experiments in the few hundred GeV to few TeV region and in
precision electroweak tests at the $Z^0$, atomic parity violation, and
polarized M{\o}ller scattering.

\bigskip

\Date{2/96}

\vfil\eject

\newsec{Introduction}

Topcolor--assisted technicolor (TC2) was proposed by Hill
\ref\tctwohill{C.~T.~Hill, Phys.~Lett.~{\bf B345}, 483 (1995).}
to overcome major shortcomings of top--condensate models of electroweak
symmetry breaking
\ref\topcref{C.~T. Hill, Phys.~Lett.~{\bf B266}, 419 (1991) \semi
S.~P.~Martin, Phys.~Rev.~{\bf D45}, 4283 (1992);
{\it ibid}~{\bf D46}, 2197 (1992); Nucl.~Phys.~{\bf B398}, 359 (1993);
M.~Lindner and D.~Ross, Nucl.~Phys.~{\bf  B370}, 30 (1992)\semi
R.~B\"{o}nisch, Phys.~Lett.~{\bf B268}, 394 (1991)\semi
C.~T.~Hill, D.~Kennedy, T.~Onogi, H.~L.~Yu, Phys.~Rev.~{\bf D47}, 2940 
(1993).},
\ref\topcondref{Y.~Nambu, in {\it New Theories in Physics}, Proceedings of
the XI International Symposium on Elementary Particle Physics, Kazimierz,
Poland, 1988, edited by Z.~Adjuk, S.~Pokorski and A.~Trautmann (World
Scientific, Singapore, 1989); Enrico Fermi Institute Report EFI~89-08
(unpublished)\semi
V.~A.~Miransky, M.~Tanabashi and K.~Yamawaki, Phys.~Lett.~{\bf
B221}, 177 (1989); Mod.~Phys.~Lett.~{\bf A4}, 1043 (1989)\semi
W.~A.~Bardeen, C.~T.~Hill and M.~Lindner, Phys.~Rev.~{\bf D41},
1647 (1990).}
and of technicolor models of dynamical electroweak and flavor symmetry
breaking
\ref\tcref{S.~Weinberg, Phys.~Rev.~{\bf D19}, 1277 (1979)\semi
L.~Susskind, Phys.~Rev.~{\bf D20}, 2619 (1979).},
\ref\etc{S.~Dimopoulos and L.~Susskind, Nucl.~Phys.~{\bf B155}, 237
(1979)\semi
E.~Eichten and K.~Lane, Phys.~Lett.~{\bf B90}, 125 (1980).}.
Technicolor and extended technicolor (ETC)
have been unable to provide a natural and plausible understanding of
why the top quark mass is so large
\ref\toprefs{F.~Abe, et al., The CDF Collaboration, Phys.~Rev.~Lett.~{\bf
73}, 225 (1994); Phys.~Rev.~{\bf D50}, 2966 (1994); Phys.~Rev.~Lett.~{\bf
74}, 2626 (1995) \semi
S.~Abachi, et al., The D\O\ Collaboration, Phys.~Rev.~Lett.~{\bf
74}, 2632 (1995).}.
On the other hand, models in which strong topcolor interactions drive
top--quark condensation {\it and} electroweak symmetry breaking are
unnatural. To reproduce the one--Higgs--doublet standard model consistent
with precision electroweak measurements (especially of the
parameter $\rho = M_W^2/M_Z^2 \cos^2\theta_W \simeq 1$), the topcolor
energy scale must be much greater than the electroweak scale of
$\CO(1\,\tev)$. This requires severe fine tuning of the topcolor
coupling.

Hill's combination of topcolor and technicolor keeps the best of both
schemes. In TC2, technicolor interactions at the scale $\Lambda_{TC} \simeq
\Lambda_{EW} \simeq 1\,\tev$ are mainly responsible for electroweak
symmetry breaking. Extended technicolor is still required for the hard
masses of all quarks and leptons {\it except} the top quark. Topcolor
produces a large top condensate, $\condtbt$, and all but a few GeV of $m_t
\simeq 175\,\gev$.\foot{A small part of $m_t$ {\it must} be generated by
ETC to give mass to the Goldstone bosons---top-pions---associated with top
condensation. Hill has pointed out that some, perhaps all, of the bottom
quark mass may arise from $\suone$ instantons~\tctwohill.} However, it
contributes comparatively little to electroweak symmetry breaking. Thus,
the topcolor scale can be lowered to near~$1\,\tev$ and the interaction
requires little or no fine tuning.

In the simplest example of Hill's TC2, there are separate color and weak
hypercharge gauge groups for the heavy third generation of quarks and
leptons and for the two light generations. The third generation transforms
under strongly--coupled $\suone \otimes \uone$ with the usual charges,
while the light generations transform conventionally under weakly--coupled
$\sutwo \otimes \utwo$. Near $1\,\tev$, these four groups are broken to the
diagonal subgroup of ordinary color and hypercharge, $\suc\otimes \uy$. The
desired pattern of condensation occurs because $\uone$ couplings are such
that the spontaneously broken $\suone \otimes \uone$ interactions are
supercritical only for the top quark.

Two important constraints were imposed on TC2 soon after Hill's proposal
was made. The first is due to Chivukula, Dobrescu and Terning (CDT)
\ref\cdt{R.~S.~Chivukula, B.~A.~Dobrescu and J.~Terning, Phys.~Lett.~{\bf
B353}, 289 (1995).}
who claimed that the technifermions required to break top and bottom quark
chiral symmetries are likely to have custodial--isospin violating couplings
to the strong $\uone$. To keep $\rho \simeq 1$, they argued,
the $\uone$ interaction must be so weak that it is necessary to fine--tune
the $\suone$ coupling to within~1\% of its critical value for top
condensation {\it and} to increase the topcolor boson mass above
$4.5\,\tev$. Thus, TC2 still seemed to be unnatural. CDT stated that their
bounds could be relaxed if $\uone$ couplings did not violate isospin.
However, they expected that this would be difficult to implement because of
the requirements of canceling gauge anomalies and of allowing mixing
between the third and first two generations.

The second constraint on TC2 is due to Kominis
\ref\kominis{D.~Kominis, Boston University Preprint BUHEP-95-20,
hep-ph/9506305 (1995).}
who showed, presuming that the $b$--quark's topcolor interactions are not
far from critical, the existence of relatively light scalar bound states of
$\ol t_L b_R$ and $\ol b_L b_R$ that couple strongly ($\propto m_t$) to
third generation quarks. These scalars can induce excessive $B_d-\ol B_d$
mixing which is proportional to the product $D^d_{Lbd} D^d_{Rbd}$ of the
elements of the unitary matrices which diagonalize the (generally
nonhermitian) $Q=-\third$ quark mass matrix.

The question of isospin violation and naturalness raised by CDT was
addressed in Ref.~\ref\tctwoklee{K.~Lane and E.~Eichten, Phys.~Lett.~{\bf
B352}, 382 (1995).}. We proposed that {\it different} technifermion
isodoublets, $T^t$ and $T^b$, give ETC mass to the top and bottom quarks.
These doublets then could have different $\uone$ charges which were,
however, isospin--conserving for the right as well as left--handed parts of
each doublet.\foot{While this eliminates the large $\rho - 1$ discussed by
CDT, there remain small, $\CO(\alpha)$, contributions from the $\uone$
interaction.} In addition, we exhibited a TC2 prototype in which (i) all
gauge anomalies cancel; (ii) there are no very light pseudo-Goldstone
bosons (loosely speaking, ``axions'') because all spontaneously broken
global technifermion symmetries are broken explicitly by
ETC~\ref\etceekl{See E.~Eichten and K.~Lane in  Ref.~\etc}; and (iii) a
mechanism exists for mixing the heavy and light generations.

Although the problem of $B_d-\ol B_d$ mixing raised by Kominis was not
considered in~\tctwoklee, the $U(1)$ symmetries of the model
presented there automatically allow just one of two ETC--induced
transitions in the quark mass matrix: $d_L,s_L \leftra b_R$ {\it or}
$d_R,s_R \leftra b_L$. Thus, only $D^d_{Lbd}$ or $D^d_{Rbd}$, respectively,
can be sizable and the $B_d-\ol B_d$ constraint is satisfied. It is easy to
see that the phenomenologically--preferred transition is $d_L,s_L \leftra
b_R$: The known mixings between the third and the first two generations are
in the Kobayashi--Maskawa matrix for left--handed quarks, $V =
D^{u\dagger}_L \ts D^d_L$. They are $|V_{cb}| \simeq |V_{ts}|
\simeq$~0.03--0.05 $\sim m_s/m_b$ and $|V_{ub}| \simeq |V_{td}| \simeq
$~0.002--0.015 $\sim \sin\theta_C \ts m_s/m_b$
\ref\pdg{Particle Data Group, Phys.~Rev.~{\bf D50}, 1174 (1994).}.
These elements must arise from $D^d_L$, hence from the $d_L,s_L \leftra
b_R$ transitions, because the corresponding elements in $D^u_L$ are smaller
by a factor of $m_b/m_t \simeq 0.03$.

In the model of Ref.~\tctwoklee, the mechanism of topcolor breaking was
left unspecified and all technifermions were taken to be
$\suone\otimes\sutwo$ singlets. Thus, the transition $d_L,s_L \leftra b_R$
had to be generated by an externally--induced term $\delta M_{ETC}$ in the
ETC mass matrix which transforms as $(\ol 3, 3)$ under the color groups. We
then estimated
\eqn\delm{|V_{cb}| \simeq |D^d_{Lsb}| \simeq {\delta m_{sb} \over {m_b}}
\simle {\delta m_{sb} \over {m_b^{ETC}}} \simeq {\delta M^2_{ETC} \over
{M^2_s}} \ts\ts,}
where $\delta m_{sb}$ is the mixing term in the $Q = -\third$ mass matrix,
$m_b$ is the mass of the $b$--quark, and $M_s$ is the mass of the ETC boson
that generates the strange--quark mass, $m_s$. In a walking technicolor
theory
\ref\wtc{B.~Holdom, Phys.~Rev.~{\bf D24}, 1441 (1981);
Phys.~Lett.~{\bf B150}, 301 (1985)\semi
T.~Appelquist, D.~Karabali and L.~C.~R. Wijewardhana,
Phys.~Rev.~Lett.~{\bf 57}, 957 (1986);
T.~Appelquist and L.~C.~R.~Wijewardhana, Phys.~Rev.~{\bf D36}, 568
(1987)\semi 
K.~Yamawaki, M.~Bando and K.~Matumoto, Phys.~Rev.~Lett.~{\bf 56}, 1335
(1986) \semi
T.~Akiba and T.~Yanagida, Phys.~Lett.~{\bf B169}, 432 (1986).},
$M_s \simge 100\,\tev$. However, we expect $\delta M_{ETC} = \CO(1\,\tev)$
because that is the scale at which topcolor breaking naturally occurs. This
gives $s$--$b$ mixing that is at least {\it 300~times} too small. We stated
in~\tctwoklee\ that providing mixing of the observed size between the heavy
and light generations is one of the great challenges to topcolor--assisted
technicolor.

This problem is addressed in the rest of this paper. I shall argue that
generational mixing is intimately connected to topcolor and electroweak
symmetry breaking and that all these phenomena occur through technifermion
condensation. In Sections~2--4, I specify the gauge groups and describe the
patterns of gauge symmetry breaking needed for standard model
phenomenology. Nontrivial patterns of vacuum alignment play a central role
in this. In Section~5, I present a class of models which illustrate this
scenario. The phenomenology of these models is sketched in Section~6.
Special attention is placed on the $Z'$ boson of the broken $\uone$
symmetry. Its effects may be noticable in hadron collider production of
jets and dileptons, $e^+e^-$ collisions, atomic parity violation, polarized
M{\o}ller scattering and other precision electroweak measurements. I also
emphasize observational consequences of vacuum alignment, especially
technirho vector mesons and their decay to pairs of technipions and,
possibly, CP~violation.

\bigskip

\newsec{Gauge Groups}

The gauge groups of immediate interest to us are
\eqn\groups{\sutc \otimes \suone \otimes \sutwo \otimes \uone \otimes
\utwo \otimes SU(2) \ts\ts,}
where, for definiteness, I have assumed that the technicolor gauge group is
$\sutc$. To avoid light ``axions'', all of these groups (except for the
electroweak $SU(2)$ and, possibly, parts of the $U(1)$'s) must be embedded
in an extended technicolor group, $\Getc$. I will not specify
$\Getc$. This difficult problem is reserved for the future. However, as in
Ref.~\tctwoklee, I shall assume the existence of ETC--induced four--fermion
operators which are needed to break quark, lepton and technifermion chiral
symmetries. Of course, these operators must be invariant under the groups
in Eq.~\groups.

The coupling constants of $\suone \otimes \sutwo \otimes \uone \otimes
\utwo$ are denoted by $g_1$, $g_2$, $g'_1$, $g'_2$, where $g_1 \gg g_2$ and
$g'_1 \gg g'_2$. When these gauge symmetries break, $\suone \otimes \sutwo
\ra \suc$ and $\uone \otimes \utwo \ra \uy$. We shall see that the breaking
to $\uy$ must occur at an energy higher than the $SU(2) \otimes
\uy$ breaking scale $\Lambda_{EW}$. Then, the usual color and weak
hypercharge couplings are
\eqn\couplings{
g_S = {g_1 g_2 \over{\sqrt{g^2_1 + g^2_2}}} \simeq g_2 \ts, \qquad
g' = {g'_1 g'_2 \over{\sqrt{g^{\prime 2}_1 + g^{\prime 2}_2}}} \simeq g'_2
\ts.}
These symmetry breakings give rise to eight color--octet ``coloron''
($V_8$) vector bosons and one neutral $Z'$, all of which have mass
of $\CO(1\,\tev)$
\ref\hp{C.~T.~Hill and S.~Parke, Phys.~Rev.~{\bf D49}, 4454 (1994)},
\tctwohill.

Third--generation quarks $q^h=(t,b)$ will transform as $(3,1)$
under $\suone\otimes \sutwo$, while the first two generation quarks $q^l =
(u,d)$, $(c,s)$ transform as $(1,3)$. Unlike the situation in the
simple models of Refs.~\tctwohill\ and \tctwoklee, we shall find it
necessary to assume that {\it all} quarks and leptons carry both $\uone$ and
$\utwo$ charges. These hypercharge assignments must be such that the
gauge interactions are supercritical only for the top quark. This new
situation has important phenomenological consequences, outlined in
Section~6.

\bigskip

\newsec{{\bf $U(1)_1 \otimes U(1)_2$ Breaking}}

In the scenario I describe, the extra $Z'$ resulting from
$\uone\otimes\utwo$ breaking has a mass of at most a few~TeV and couples
strongly to light, as well as heavy, quarks and leptons. Then, two
conditions are necessary to prevent conflict with neutral current
experiments. First, there must be a $Z^0$~boson with standard electroweak
couplings to all quarks and leptons. To arrange this, there will be a
hierarchy of symmetry breaking scales, with $\uone\otimes\utwo \ra \uy$ at
1--2~TeV, followed by $SU(2) \otimes \uy \ra U(1)_{EM}$ at the lower scale
$\Lambda_{EW}$. Assuming that technicolor interactions induce both symmetry
breakdowns, the technifermions responsible for $\uone\otimes\utwo \ra
\uy$---call them $\psi_L$ and $\psi_R$---must belong to a {\it vectorial}
representation of $SU(2)$. To simplify the analysis, I make the minimal
assumption that the $\psi_{L,R}$ are electrically neutral $SU(2)$ singlets.

To produce this hierarchy of symmetry breaking scales, and yet maintain an
asymptotically free technicolor, the $\psi_{L,R}$ should belong to a {\it
higher--dimensional} representation of $\sutc$, while the technifermions
responsible for $SU(2)\otimes \uy$ breaking must belong to fundamental
representations. This is reminiscent of multiscale technicolor
\ref\multi{K.~Lane and E.~Eichten, Phys.~Lett.~{\bf B222}, 274 (1989) \semi
K.~Lane and M~V.~Ramana, Phys.~Rev.~{\bf D44}, 2678 (1991).},
but there both the higher and fundamental representations participate in
electroweak symmetry breaking. In the present model, I shall assume that
$\psi_{L,R}$ belong to the $\half \Ntc (\Ntc - 1)$--dimensional
antisymmetric tensor representation. I assume that this set of
technifermions is large enough to ensure that the technicolor coupling
``walks'' for a large range of momenta~\wtc.

The second constraint is that the $Z'$ should not induce large
flavor--changing interactions. This can be achieved if the $\uone$
couplings of the two light generations are GIM--symmetric. Then
flavor--changing effects will nominally be of order $|V_{ub}|^2/M^2_{Z'}$
for $\Delta B_d = 2$ processes, $|V_{cb}|^2/M^2_{Z'}$ for $\Delta B_s = 2$,
and negligibly small for $\Delta S = 2$. These should be within
experimental limits.\foot{The most stringent constraint may come from
$\Delta M_{B_d}/M_{B_d}$. In the model of Section~5, this ratio depends in
a complicated way on the $\uone$ hypercharges $b$, $b'$, $d$, $d'$ and the
magnitudes and phases of $V_{ub}$ and $V_{td}$.} Nevertheless, a variety of
interesting, and potentially dangerous, $Z'$~phenomena are expected. These
are discussed in Section~6.

\bigskip

\newsec{{\bf $SU(3)_1\otimes SU(3)_2$} and Electroweak Breaking and
Generational Mixing}

Turn now to symmetry breaking at lower energy scales. I recounted above
that $s$--$b$ mixing is too small by a factor of~300 if $\suone \otimes
\sutwo$ breaking is introduced to the quark sector only by a mixing term in
the ETC boson mass matrix. Since $b_R$ transforms as $(3, 1, 1; -\third)$
under $\suone \otimes \sutwo \otimes SU(2) \otimes \uy$ and $d_L, s_L$ as
$(1, 3, 2; \sixth)$, it is tempting to suppose that the mechanism
connecting $d_L,s_L$ to $b_R$ is at the same time responsible for breaking
$\suone \otimes \sutwo \ra \suc$ {\it and} $SU(2)\otimes \uy \ra
U(1)_{EM}$. The generational mixing term transforms as $(\ol 3, 3)$ under
the color groups. Therefore, I introduce colored technifermion isodoublets
transforming under $\sutc \otimes \suone \otimes \sutwo \otimes SU(2)$ as
follows:
\eqn\tquarks{\eqalign{
&T^1_{L(R)} = \pmatrix{U^1 \cr D^1\cr}_{L(R)} \in (\Ntc, 3, 1, 2(1)) \cr
&T^2_{L(R)} = \pmatrix{U^2 \cr D^2\cr}_{L(R)} \in (\Ntc,1,3, 2(1)) \ts\ts.
\cr}}
The transition $d_L, s_L \leftra D^2_L \leftra D^1_R \leftra b_R$
occurs if the appropriate ETC operator exists {\it and} if the condensate
$\condab$ forms.

The patterns of condensation, $\condij$, that occur depend on the strength
of the interactions driving them and on explicit chiral
symmetry breaking (4T)~interactions that determine the correct
chiral--perturbative ground state, i.e., ``align the vacuum''
\ref\vacalign{R.~Dashen, Phys.~Rev.~{\bf D3}, 1879 (1971);
S.Weinberg, Phys.~Rev.~{\bf D13}, 974 (1976)\semi
E.~Eichten, K.~Lane and J.~Preskill, Phys.~Rev.~Lett.~{\bf 45}, 225
(1980)\semi
K.~Lane, Physica Scripta~{\bf 23}, 1005 (1981)\semi
M.~Peskin, Nucl~Phys.~{\bf B175}, 197 (1980);
J.~Preskill, Nucl.~Phys.~{\bf B177}, 21 (1981).}.
The strong interactions driving technifermion condensation are $\sutc$,
$\suone$ and $\uone$. The technicolor interactions do not prefer any
particular form for $\condij$; $\suone$ drives $\condaa \neq 0$; $\uone$
drives $\condaa, \ts\ts \condbb \neq 0$ {\it or} $\condab \neq 0$,
depending on the strong hypercharge assignments.

In the approximation that technicolor interactions dominate condensate
formation, so that
\eqn\cij{\condij = -\half \Delta_T U_{ij} \qquad (i,j = 1,2) \ts,}
it is easy to prove the following: If $T^1 \in (3, 1)$ and $T^2 \in (1, 3)$
are the only technifermions and if the vacuum--aligning interactions are
$\suone \otimes \sutwo$ symmetric then, in each charge sector, the unitary
matrix $U_{ij} = \delta_{ij}$ or $U_{ij} = (i\sigma_2)_{ij}$, but {\it not}
a nontrivial combination of the two. Therefore, in order that $\suone
\otimes \sutwo$--invariant direct mass terms, $d_L,s_L \leftra d_R,s_R$ and
$b_L \leftra b_R$, occur as well as the mixing $d_L, s_L \leftra b_R$, it
is necessary to introduce still other technifermions. The least number of
additional technifermions involves $\suone \otimes \sutwo$ singlets. In the
model described below, these will consist of three isodoublets: $T^l$
giving direct mass terms to the light quarks and leptons; $T^t$ giving the
top quark its ETC mass; and $T^b$ giving the bottom quark its ETC mass.
These are the same technifermions used in the model of Ref.~\tctwoklee.
Introducing them enlarges the chiral symmetry---and the number of Goldstone
bosons---of the model. Giving mass to all these bosons will require, among
other things, a nontrivial pattern of $T^1$--$T^2$ condensation, $U = a_0 1
+ i a_2 \sigma_2$. This simultaneously breaks the color and electroweak
symmetries to $\suc\otimes U(1)_{EM}$ and provides large generational
mixing, e.g., $\delta m_{sb} \sim \langle \ol T^1 T^2 \rangle_{M_s}/M^2_s
\sim m_s$. The color--singlet technifermions help align the vacuum in this
nontrivial way as well as contribute to electroweak symmetry breaking.

\bigskip

\newsec{A Model}

In this section I follow the format of Ref.~\tctwoklee\ to construct a TC2
model with the symmetry breaking just outlined. First, I list hypercharge
assignments for all the fermions and explain certain general constraints on
them. Then I derive a condition on the hypercharges that must be satisfied
in order that colored technifermions condense to break topcolor~$SU(3)$. I
conclude by discussing other conditions available to fix the hypercharges.
Among these are the gauge anomaly constraints, given in the Appendix. The
rest follow from specifying the ETC four--fermion operators necessary to
give masses to quarks and leptons and to the Goldstone bosons associated
with global symmetries. A family of solutions for the hypercharges
satisfying all these constraints is obtained in the Appendix.

The fermions in the model, their color representations and $U(1)$ charges
are listed in Table~1. A number of choices have been made at the outset to
limit and simplify the charges and to achieve the scenario's objectives:

\medskip

\item{1.} In order that technifermion condensates conserve electric charge,
$u_1 + u_2 = v_1 + v_2$, $x_1 + x_2 = x'_1 + x'_2$, $y_1 + y_2 = y'_1 +
y'_2$, and $z_1 + z_2 = z'_1 + z'_2$.

\item{2.} The $\uone$ charges of technifermions respect custodial isospin.

\item{3.} The most important choice for our scenario is that of the $\uone$
charges of $T^1$ and $T^2$. So long as $u_1 \neq v_1$, the broken $\uone$
interactions favor condensation of $T^1$ with $T^2$. If this interaction is
stronger than the $\suone$--attraction for $T^1$ with itself and if we
neglect other vacuum--aligning ETC interactions, then $\condij \propto
(i \sigma_2)_{ij}$ in each charge sector. This alignment is discussed below.

\item{4.} We shall see that $u_1 \neq v_1$ implies $Y_{1i} \neq Y'_{1i}$
for the various fermions. Purely for simplicity, I have chosen $Y_1 = b'$
for all right--handed light quarks. I must choose $Y_1(t_R) \neq Y_1(b_R)$
to prevent strong $b$--condensation. Again for simplicity, I put $Y_1(t_R)
= -Y_1(b_R) = d'$. We shall see that $dd'$ is positive, as it must be for
$t$--condensation.

\item{5.} For the $\sutc$ antisymmetric tensor $\psi$, $\xi' \neq \xi$
guarantees $\uone \otimes \utwo \ra \uy$ when $\langle \ol \psi_L \psi_R
\rangle$ forms. Note that, if $\Ntc = 4$, a single real $\psi_L$ is
sufficient to break the $U(1)$'s. Otherwise, to limit the parameters, $\xi'
= -\xi$ may be assumed.

\medskip

I now show that, in the absence of other ETC operators, the $\uone$
interactions can overwhelm $\suone$ to produce the alignment pattern
$\condij \propto (i \sigma_2)_{ij}$. The coupling of the $Z'$ boson to a
generic fermion $\chi$ with weak hypercharge $Y=Y_1 + Y_2$ and electric
charge~$Q = Y'_1 + Y'_2$ is
\eqn\LxZx{\CL_{\ol \chi Z' \chi} = g_{Z'} Z^{\prime\mu}
\left[\ts \left(Y_1 - rY \right) \ts \ol \chi_L \gamma_\mu \chi_L +
\left(Y^\prime_1 - rQ \right) \ts \ol \chi_R \gamma_\mu \chi_R \right]
\ts\ts,}
where $g_{Z'} = \sqrt{g^{\prime 2}_1 + g^{\prime 2}_2} \simeq g_1'$ and $r
= g^{\prime 2}_2/g_{Z'}^2 \ll 1$. Small mixing terms induced by electroweak
symmetry breaking have been neglected in this expression. A similar
interaction can be written for the massive $V_8$ bosons of broken $\suone
\otimes \sutwo$. Ignoring small terms in the $Z'$ and $V_8$ couplings, the
four--fermion interaction these bosons generate for $T^1$ and $T^2$ is
\eqn\LTT{\eqalign{
\CL_{T^1 T^2} = & \ts\ts
-2\pi \biggl\{ {\alpha_{Z'} \over {M^2_{Z'}}} \biggl[
u_1 \bigl(\ol T^1_L \gamma_\mu T^1_L + \ol T^2_R \gamma_\mu T^2_R
\bigr) + v_1 \bigl(\ol T^1_R \gamma_\mu T^1_R + \ol T^2_L \gamma_\mu
T^2_L\bigr) \ts \biggr]^2_{M_{Z'}}\cr
&\qquad +{\alpha_{V_8} \over {M^2_{V_8}}}
\sum_a \bigl(\ol T^1_L \gamma_\mu t_a T^1_L + \ol T^1_R \gamma_\mu t_a
T^1_R \bigr)^2_{M_{V_8}} \biggr\} \ts\ts, \cr}}
where $\alpha_{Z'} = g^2_{Z'}/4\pi$ and the $t_a$ are $SU(3)$ matrices in
the $3$-representation. All the currents are $\sutc \otimes SU(2)$
singlets, and the current $\times$ current products are renormalized at the
corresponding massive boson masses. Fierzing this interaction and retaining
only the dominant $SU(3) \otimes \sutc \otimes SU(2)$--singlet operators
involved in condensate formation gives
\eqn\fierz{\eqalign{
\CL_{T^1 T^2} = & {4\pi \over {3 \Ntc M^2_{V_8}}}
\biggl[{u_1 v_1 \alpha_{Z'} M^2_{V_8} \over {M^2_{Z'}}}
\bigl(\ol T^1_L T^1_R \ol T^1_R T^1_L  + \ol T^2_L T^2_R \ol T^2_R T^2_L 
\bigr)_{M_{Z'}} \cr
& \qquad\qquad  + {\alpha_{Z'} M^2_{V_8} \over {M^2_{Z'}}}
\bigl(u^2_1 \ts \ol T^1_L T^2_R \ol T^2_R T^1_L  + v^2_1 \ts \ol T^2_L
T^1_R \ol T^1_R T^2_L \bigr)_{M_{Z'}} \cr
&\qquad\qquad + {4 \alpha_{V_8} \over {3}}
\bigl(\ol T^1_L T^1_R \ol T^1_R T^1_L \bigr)_{M_{V_8}} \ts \biggr]
\ts\ts. \cr}}

To determine which of the operators in Eq.~\fierz\ is dominant, I make
the large--$\Ntc$ approximation that the anomalous dimensions of the
4T~operators are given by the sum of the anomalous dimensions
$\gamma_{m_{ij}}$ of their constituent bilinears $\ol T^i T^j$. Then, the
condition that the vacuum energy $E = - \langle \CL_{T^1 T^2} \rangle$
is minimized by $\condij \propto (i \sigma_2)_{ij}$ is 
\eqn\sigtwo{\eqalign{
&{\alpha_{Z'} (u^2_1 + v^2_1) M^2_{V_8} \over {M^2_{Z'}}}
\ts {Z^2_{12}(M_{Z'}) \over {Z^2_{11}(M_{V_8})}} \cr\cr
&\quad > {4 \alpha_{V_8} \over {3}} +{u_1 v_1 \alpha_{Z'} M^2_{V_8}
\over {M^2_{Z'}}} \ts {Z^2_{11}(M_{Z'}) + Z^2_{22}(M_{Z'}) \over
{Z^2_{11}(M_{V_8})}} \ts\ts, \cr}}
where
\eqn\Zij{Z_{ij}(M) = \exp\left[\int_{\Lambda_{TC}}^M {d\mu  \over{\mu}}
\gamma_{m_{ij}}(\mu) \right] \ts\ts.}
Since the $U(1)$ symmetries are broken at a higher scale than the $SU(3)$
and electroweak symmetries, $M_{Z'}$ may be several times larger
than $M_{V_8}$. However, the energy range from $M_{V_8}$ to $M_{Z'}$
overlaps the region in which $T$--condensates form.
Thus, the anomalous dimensions $\gamma_{m_{ij}} \simeq 1$ there~\wtc. In
this limit, the condition \sigtwo\ becomes $(u_1 - v_1)^2 >
4\alpha_{V_8}/3\alpha_{Z'}$.

The rest of my discussion of this model concerns how the $\uone$ and
$\utwo$ hypercharges are to be fixed. I start with the gauge anomaly
conditions. The eight independent conditions are given in the Appendix.
These constraints, together with the 4~equal--charge conditions, do not fix
the 26 unknown $U(1)_i$ charges. Further limitations on the $Y_i$ follow
from requiring the presence of ETC--generated four--fermion operators
breaking all but gauged symmetries. To give mass to quarks and leptons, I
assume the following ETC operators:
\eqn\qTTq{\eqalign{
\ol \ell^l_{iL} \gamma^\mu T^l_L \ts \ol D^l_R \gamma_\mu e_{jR}
\qquad &\Lra \qquad a - a' = x_1 - x'_1 \cr
\ol q^l_{iL} \gamma^\mu T^l_L \ts \ol T^l_R \gamma_\mu q^l_{jR} 
\qquad &\Lra \qquad b - b' = x_1 - x'_1 \cr
\ol \ell^h_L \gamma^\mu T^l_L \ts \ol D^l_R \gamma_\mu \tau_R
\qquad &\Lra \qquad c - c' = x_1 - x'_1 \cr
\ol q^h_L \gamma^\mu T^t_L \ts \ol U^t_R \gamma_\mu t_R
\qquad &\Lra \qquad d - d' = y_1 - y'_1 \cr
\ol q^h_L \gamma^\mu T^b_L \ts \ol D^b_R \gamma_\mu b_R
\qquad &\Lra\qquad d + d' = z_1 - z'_1 \ts\ts. \cr}}
To generate $d_L,s_L \leftra b_R$, I require the operator
\eqn\sLbR{\ol q^l_{iL} \gamma^\mu T^2_L \ts \ol D^1_R \gamma_\mu b_R
\qquad \Lra \qquad b + d' = 0 \ts\ts.}
To forbid $d_R,s_R \leftra b_L$, ETC interactions must not generate the
operator  $\ol q^h_L \gamma^\mu T^1_L \ts \ol D^2_R \gamma_\mu d_{iR}$.
This gives the constraint
\eqn\dbp{ d-b' \neq 0 \ts.}
We shall see that this follows from requiring the existence of other
four--fermion operators and also the anomaly constraints. Thus, this
operator does not appear without the intervention of $\uone$ breaking and
so the transition $d_R,s_R \leftra b_L$ is automatically suppressed
relative to $d_L,s_L \leftra b_R$ by a factor of $\delta M^2_{ETC}/M^2_s =
\CO(10^{-4})$.

Next, I enumerate the chiral symmetries and Goldstone bosons of the model,
to determine what 4T operators are needed to give them mass. The simplest
way to do this is to imagine that all gauge interactions, including $\suone
\otimes \sutwo \otimes \uone$, may be neglected compared to technicolor.
Then, grouping the technifermions into three triplet--isodoublets, $T^1,
T^2$ and $T^3 = T^l, T^t, T^b$, the chiral symmetry group of these
technifermions plus $\psi_{L,R}$ is
\eqn\Gchi{G_\chi = SU(18)_L \otimes SU(18)_R \otimes U(1)_A \ts.}
The $U(1)_A$ current involves all technifermions and has no technicolor
anomaly. It is spontaneously broken principally by $\langle \ol \psi_L
\psi_R \rangle$. A linear combination of this current and generators of
$SU(18)_A$ is exactly conserved and couples to the Goldstone boson eaten by
the $Z'$. The orthogonal Goldstone boson gets mass from $\suone$ instantons
and broken ETC interactions. We need not be further concerned with
$U(1)_A$.

When $T$--condensates break $SU(18)_L \otimes SU(18)_R$ to an $SU(18)$
subgroup, there are 323 Goldstone bosons or technipions, $\tpi$.\foot{I do
not know whether this is a record number of Goldstone bosons, as has been
speculated. It certainly is a matter of concern whether they may make a
large positive contribution to the $S$--parameter. This is the case if they
may be approximated as pseudo-Goldstone bosons~\ref\pettests{
A.~Longhitano, Phys.~Rev.~{\bf D22} (1980) 1166; Nucl.~Phys.~{\bf B188},
(1981) 118\semi R.~Renken and M.~Peskin, Nucl.~Phys.~{\bf B211}
(1983) 93\semi
B.~W.~Lynn, M.~E.~Peskin and R~.G.~Stuart, in Trieste Electroweak 1985,
213\semi
M.~Golden and L.Randall, Nucl.~Phys.~{\bf B361} (1990) 3\semi B.~Holdom and
J.~Terning, Phys.~Lett.~{\bf B247} (1990) 88\semi
M.~E.~Peskin and T.~Takeuchi, Phys.~Rev.~Lett.~{\bf 65} (1990) 964\semi
A.~Dobado, D.~Espriu and M.~J.~Herrero, Phys.~Lett.~{\bf B255}
(1990) 405\semi
H.~Georgi, Nucl.~Phys.~{\bf B363} (1991) 301.}. As I have discussed
elsewhere~\ref\ichep{K.~Lane, Proceedings of the 27th International
Conference on High Energy Physics, edited by P.~J.~Bussey and
I.~G.~Knowles, Vol.~II, p.~543, Glasgow, June 20--27, 1994.}, this may be a
poor approximation for the technipions in a walking technicolor model with
its large anomalous dimensions. Furthermore, in such a model, there are
additional, possibly negative, contributions to~$S$ which cannot be
evaluated simply by scaling from QCD (see also Ref.~\ref\tajtstu{B.~Holdom,
Phys.~Lett.~{\bf B259}, 329 (1991)\semi E.~Gates and J.~Terning,
Phys.~Rev.~Lett.~{\bf 67}, 1840 (1991)\semi M.~Luty and R.~Sundrum,
Phys.~Rev.~Lett.~{\bf 70}, 127 (1993)\semi T.~Appelquist and J.~Terning,
Phys.~Lett.~{\bf B315}, 139 (1993).}).} These may be conveniently
classified according to the subgroup
\eqn\Hchi{H_\chi = \suone \otimes \sutwo \otimes \suthree
\otimes SU(2) \otimes U(1)_3 \otimes U(1)_8 \ts,}
where $SU(3)_i$ acts on the triplet $T^i$, $SU(2)$ acts on the isodoublets
within the triplets, and $U(1)_{3,8}$ are generated by the diagonal charges
of the $SU(3)$ defined on the triplet $T^1,T^2,T^3$:
\eqn\Treps{T^1 \in (3, 1, 1, 2; \half, \sqrt{\twelveth})\ts\ts, \quad  T^2
\in (1, 3, 1, 2; -\half, \sqrt{\twelveth})\ts\ts, \quad  T^3 \in (1, 1, 3,
2; 0, -\sqrt{\third}) \ts\ts.}
The 323 Goldstone bosons consist of: three $SU(3)$--singlet isotriplets,
$(1,1,1,3)$; three octet isotriplets plus three octet isosinglets; two
singlets, $(1,1,1,1)$; and three sets of $(3,\ol 3) \oplus (\ol 3, 3)$
isotriplets and isosinglets.

The diagonal linear combination of the three $(1,1,1,3)$'s become $W^\pm_L$
and $Z^0_L$. Thus, ignoring the effects of color interactions, the decay
constant of the technipions is $F_T = 246\,\gev/\sqrt{9} =
82\,\gev$.\foot{I am suppressing the role of the $SU(2) \otimes U(1)$
chiral symmetry of $(t,b)_L$ and $t_R$ in this discussion. The three
Goldstone top-pions, $\toppi$, arising from its breakdown combine with the
$(1,1,1,3)$'s to form the longitudinal weak bosons. In our normalization,
Hill's estimate of the top-pion decay constant is $F_t \simeq
35\,\gev$~\tctwohill. The uneaten component of the top-pions acquires its
mass from the ETC part of the top quark mass: $M^2_{\pi_t} \simeq m^{ETC}_t
\condtbt/F^2_t$.} A linear combination of the $(8,1,1,1)$ and $(1,8,1,1)$
are absorbed in $\suone \otimes \sutwo \ra \suc$, driven by $\condab$. Of
the remaining 312 Goldstone bosons, all those which are $\suone \otimes
\sutwo$ nonsinglets (there are 272 of these) acquire mass of at least
$\sqrt{(\alpha_S \Lambda^4_T/F^2_T)} \simeq 250\,\gev$ from color
interactions (see the papers by Peskin and Preskill in Ref.~\vacalign).

This leaves 40 technipions whose mass must arise from ETC--generated
4T~interactions. They transform as $(1,1,8,3) \oplus (1,1,8,1) \oplus
(1,1,1,3) \oplus (1,1,1,3) \oplus (1,1,1,1) \oplus (1,1,1,1)$. Consider the
two isotriplets $(1,1,1,3)$ orthogonal to the longitudinal weak bosons. It
is possible to form one linear combination of these states that contains no
$\ol T^i T^i$ component for one of the values of $i=1,2,3$. Therefore,
there must be a 4T~term involving two technifermions of the form $\ol T^i_L
\gamma^\mu T^j_L \ol T^j_R \gamma_\mu T^i_R$, with $i \neq j$, to insure
that both isotriplets get mass. The only operators consistent with $u_1 -
v_1 \neq 0$ have $i = 1$ or~2 and $j=3$, with $T^3 = T^l$ or $T^t$ or
$T^b$. Finally, in order that such an interaction contribute to $(1,1,1,3)$
technipion masses, it is necessary that the condensates $\condaa$ and
$\condbb$ form, i.e., that the matrix $U$ in Eq.~\cij\ is a nontrivial
combination of~$1$ and $i \sigma_2$. Any of these 4T~operators, in concert
with $\sutc \otimes \suone \otimes \uone$ interactions, can lead to such a
pattern of vacuum alignment. As a specific choice consistent with
Eqs.~\qTTq\ and \sLbR, I assume the existence of the operator
\eqn\Htt{\ol T^1_L \gamma^\mu T^t_L \ts \ol T^t_R \gamma_\mu (a + b
\sigma_3) T^1_R \qquad \Lra \qquad y_1 - y'_1 = u_1 - v_1 \ts.}

Equations~\qTTq, \sLbR, \Htt\ and the anomaly conditions for
$U(1)_{1,2} [\sutc]^2$ and  $U(1)_{1,2} [SU(3)_{1,2}]^2$ lead to the
relations:
\eqn\charges{\eqalign{
&a - a' = b - b' = c - c' = x_1 - x'_1 = \half\Ntc(u_1 - v_1) \cr
&d - d' = y_1 - y'_1 = u_1 - v_1 \cr
&d + d' = z_1 - z'_1 = -(2\Ntc + 1)(u_1 - v_1) \cr
&d = -\Ntc(u_1-v_1) \cr
&d' = -b = -(\Ntc + 1)(u_1 - v_1) \cr
&b' = \half(\Ntc + 2)(u_1 -v_1) \cr
& (\Ntc -2)(\xi - \xi') = 3\Ntc (u_1 - v_1) \ts\ts. \cr}}
We see that the constraint $d-b'\neq 0$ forbidding $d_R,s_R \leftra s_L$ is
satisfied. Also, $dd' > 0$, just what is needed for top, but not bottom,
quarks to condense. The condition $\xi-\xi' \neq 0$ that $\langle \ol
\psi_L \psi_R \rangle$ breaks $\uone\otimes\utwo \ra \uy$ is equivalent
to $u_1 - v_1 \neq 0$, necessary for $\condab \neq 0$.

Finally, there are $(1,1,8,3)$, $(1,1,8,1)$ and $(1,1,1,1)$ technipions
composed of $T^l$ and $T^b$ that do not acquire mass from the operator in
Eq.~\Htt. Combinations of spontaneously broken currents such as $\ol T^2
\gamma_\mu \gamma_5 T^2 - 3 \ol T^b \gamma_\mu \gamma_5 T^b$ are also left
conserved by this operator. Thus, we need a 4T~operator involving both $T^l$
and $T^b$. One choice (of several) that is consistent with all the
operators assumed so far is
\eqn\Hltbl{\ol T^l_L \gamma^\mu T^t_L \ts \ol T^b_R \gamma_\mu (a + b
\sigma_3) T^l_R \quad \Lra \quad y_1 - z'_1 = z'_2 - y_2
= x_1 - x'_1 = \half\Ntc(u_1 - v_1) \ts.}
Note that $T^t$ and $T^b$ must have the same electric charges, i.e., $y_1 +
y_2 = z_1 + z_2$.

We now have 18~linear plus 3~nonlinear conditions on the 26~hypercharges.
In the Appendix, I exhibit solutions to these equations for which $|u_1 -
v_1| = \CO(1)$. The vacuum alignment program, including determination of
the eigenvalues and eigenstates of the technipion mass matrix, is outlined
in Section~6 and then deferred to a later paper.

\bigskip

\newsec{The Phenomenology of Topcolor--Assisted Technicolor}

The picture of topcolor--assisted technicolor I have drawn in this paper
leads to a wide variety of phenomena in the TeV~energy region, many of
which are likely to be accessible in Tevatron collider experiments and,
possibly, in LEP2 experiments. Here is a list of the more obvious issues:

\medskip

\item{1.} The $Z'$ boson, with $M_{Z'} = 1$--$3\,\tev$.

\item{2.} The $V_8$ colorons, with mass $M_{V_8} \simle 1\,\tev$. Their
phenomenology was discussed in Refs.~\hp\ and~\ref\kltop{K.~Lane,
Phys.~Rev.~{\bf D52}, 1546 (1995).}.

\item{3.} The quantum numbers, masses, and production and decay modes of
technirhos, technipions and top-pions.

\item{4.} A possible outcome of vacuum alignment is the appearance of
CP--violating phases in the unitary matrices defining mass eigenstate
quarks (see Eichten, Lane and Preskill in Ref.~\vacalign).

\item{5.} Cosmological consequences of the $\psi$ fermion which, apparently,
must have a component that is stable against weak decay.

\item{6.} Since $|u_1 - v_1|$ must be $\CO(1)$, some of the hypercharges in
Eq.~\charges\ are $\CO(\Ntc)$. This raises the question of the triviality
of the $\uone$ interaction: does it set in at an energy much
lower than the one at which we can envisage $\uone$ being unified into an
asymptotically free ETC group?

\medskip

\noindent Each of these topics requires extensive study. Here, I briefly
discuss only the $Z'$ and the aspects of vacuum alignment. Details are
under investigation by others or postponed to later papers

\medskip

\noindent {\it {\underbar {$Z'$ Physics}}}

The mass of the $Z'$ arises mainly from $\psi$--condensation,
\eqn\mzp{M_{Z'} \simeq g_{Z'} \ts |\xi - \xi'| \ts F_\psi \ts,}
where $\xi - \xi' = 3\Ntc(u_1 - v_1)/(\Ntc - 2) = \CO(1)$, and $F_\psi =
\CO(1\,\tev)$ is the $\pi_\psi$ decay constant. This is the basis of my
estimate of $M_{Z'}$. The $Z'$ decays into technifermion, quark and lepton
pairs, with large couplings to all. Thus, its width is large, probably
several hundred~GeV~\hp. I emphasize that in this scenario the $Z'$
necessarily couples strongly to the first two generations of quarks and
leptons.

There are several precision electroweak studies that probe for the $Z'$
\ref\bu{Studies along these lines are in progress by R.~S.~Chivukula
and J.~Terning.}.
Mixing of the $Z'$ and $Z^0$ affects the latter's couplings to quark
and lepton pairs. If the $Z'$ width is not an issue, the magnitude of these
mixing effects is
\eqn\thzzp{\theta_{ZZ'} \simeq { g_{Z'} M^2_{Z} \over {g_Z M^2_{Z'}}} \ts,}
where $g_Z = \sqrt{g^2 + g^{\prime 2}}$. This mixing also affects the
$S$--parameter~\pettests.

Mixing and direct $Z'$ interactions together influence other, very 
low--energy measurements. For example, in the class of models outlined
above, the electron has an axial-vector coupling to the $Z'$. This is
probed in atomic parity violation experiments, which are especially
sensitive to the product of this coupling with the vector part of the
isoscalar nuclear current
\ref\pgl{P.~Langacker,{\it Theoretical Study of the Electroweak
Interaction---Present and Future}, Proceedings of the 22$^{\rm nd}$~INS
Symposium on Physics with High Energy Colliders, Tokyo, March 1994.}.
The effective interaction is
\eqn\Lapv{
\CL_{{\rm APV}} = -{g^2_{Z'} (a'-a) (b'+b) \over {4 M^2_{Z'}}}\ts
\ol e \gamma^\mu \gamma_5 e \ts (\ol u \gamma_\mu u  + \ol d \gamma_\mu d)
\ts.}
The product $(a'-a)(b'+b)/4 = -\Ntc (3 \Ntc + 4)(u_1 - v_1)^2/16$
can be large in this model. Out of concern for this, I have tried to
construct models within the present framework in which the the electron's
coupling to $Z'$ is purely vectorial. So far, I have not found one that has
a nontrivial ($u_1 - v_1 \neq 0$) solution to the anomaly conditions.

As a second example, the polarized M{\o}ller scattering experiment recently
proposed by Kumar and his collaborators
\ref\kumar{K.~Kumar, E.~Hughes, R.~Holmes and P.~Souder, ``Precision Low
Energy Weak Neutral Current Experiments'', Princeton University
(October 30, 1995), to appear in Mod.~Phys.~Lett.~A.}
is sensitive to the combination $a^{\prime 2} - a^2$ of electron couplings
to the $Z'$. The effective interaction is (apart from mixing effects)
\eqn\Lmoller{
\CL_{{\rm Moller}} = -{g^2_{Z'}\over {2 M^2_{Z'}}}\ts
\biggl[a^2 (\ol e_L \gamma_\mu e_L)^2 +
a^{\prime 2} (\ol e_R \gamma_\mu e_R)^2 \ts\biggr]
\ts.}

The $Z'$ will also be visible in {\it current} and planned high--energy
collider experiments. At subprocess energies well below the $Z'$ mass, its
effects are still well--approximated by four--fermion ``contact''
interactions, similar to those expected for composite quarks and leptons
\ref\elp{E.~Eichten, K.~Lane and M.~Peskin, Phys.~Rev.~Lett.~{\bf 50}, 811
(1983).}.
Thus, at the Tevatron collider, the $Z'$s strong couplings to quarks
produce an excess of high--$E_T$ jets\foot{The $V_8$ colorons enhance only
$t \ol t$ and $b \ol b$ production.}$^{,}$\foot{As this paper was being
completed, I received two preprints discussing the possibility that a
TeV--mass $Z'$ boson affects high--$E_T$ jet production and the
branching ratios for $Z^0$ decay to $\ol b b$ and $\ol c
c$~\ref\zprime{P.~Chiappetta, et al., CPT-96/P.3304, hep-ph~9601306\semi
G.~Altarelli, et al., CERN-TH/96-20, hep-ph~9601324.}.} and high--mass
dileptons. The effective interactions are
\eqn\Ltev{\eqalign{
\CL_{qq} = &-{g^2_{Z'}\over {2 M^2_{Z'}}}\ts
\biggl[\sum_{q = u,d,c,s} (b \ts \ol q_L \gamma_\mu q_L + b' \ts \ol q_R
\gamma_\mu q_R)  \cr &\qquad\qquad
+ d \ts (\ol t_L \gamma_\mu t_L + \ol b_L
\gamma_\mu b_L) + d'\ts (\ol t_R \gamma_\mu t_R - \ol b_R
\gamma_\mu b_R) \ts \biggr]^2 \ts; \cr\cr
\CL_{q\ell} = &-{g^2_{Z'}\over {M^2_{Z'}}}\ts
\sum_{q = u,d,c,s} (b \ts \ol q_L \gamma^\mu q_L + b' \ts \ol q_R
\gamma^\mu q_R)  \ts \sum_{\ell = e,\mu} (a \ts \ol \ell_L \gamma_\mu
\ell_L + a' \ts \ol \ell_R \gamma_\mu \ell_R)  \ts. \cr}}
In these expressions, we have ignored small effects of mixing among quark
generations. Note that there are simplifications of the couplings such as
$g^2_{Z'} b^2/M^2_{Z'} \simeq [(\Ntc+1)(\Ntc-2)/3\Ntc F_\psi]^2$.
The $Z'$ interaction affecting Bhabha scattering and muon--pair
production in $e^+e^-$ collisions is
\eqn\Lll{\CL_{\ell\ell} = -{g^2_{Z'}\over {2 M^2_{Z'}}}\ts
\biggl[\sum_{\ell = e,\mu} (a \ts \ol \ell_L \gamma_\mu \ell_L +
a' \ts \ol \ell_R \gamma_\mu \ell_R) \ts \biggr]^2 \ts.}
Jet production in $e^+e^-$ collisions is modified by $\CL_{ql}$.
Corresponding interactions influence tau--pair production. At the LHC, the
excess of high--$E_T$ jets will be enormous and the $Z'$ shape should be
observable as a resonance in dileptons if not in dijets. A high luminosity
$e^+e^-$ collider with $\ecm \simeq M_{Z'}$ can make detailed studies of
the $Z'$ couplings. One with $\ecm \simeq 500\,\gev$ may be able to detect
signs of $\gamma$--$Z$--$Z'$ interference.

\medskip
\noindent {\it {\underbar {Vacuum Alignment and Technihadron Physics}}}

The spectrum of technirhos $\trho$ in this model is the same as that given
above for the technipions. Determining the mass--eigenstate $\tpi$ and
$\trho$ is the problem of vacuum alignment in the technifermion sector.
This is essentially the same as diagonalizing the technifermion mass matrix
(see, however, footnote~4 for a caveat on the use of chiral perturbation
theory.) The top-pions $\toppi$ formed from $(t,b)_L$
and $t_R$ must be added to this large $\tpi$--diagonalization calculation.
Once mass eigenstates are determined, the $\trho \ra \tpi \tpi$ couplings
can be determined by symmetry (see, e.g., \multi). Note that the $\trho$
decay modes may include one or two weak bosons, $W^\pm_L$ and $Z^0_L$.
Vacuum alignment also determines the pattern of technifermion condensation,
relevant for mixing between heavy and light quarks, and feeds into the
Kobayashi--Maskawa matrix and other quark mixing angles and phases.

Vacuum alignment is carried out by minimizing the ground--state energy of
broken--ETC and $\suone \otimes \uone$ four--fermion operators and of
second--order QCD interactions~\vacalign. In the absence of a concrete ETC
model, the most that can be done is to make ``reasonable'' guesses for the
coefficients of allowed operators---those already assumed plus others
consistent with symmetries. Different assumptions for the relative
strengths and signs of the operators will lead to different vacua, patterns
of condensation, and $\tpi$ and $\trho$ spectroscopies. Such studies should
give us a plausible range of expectations for this aspect of TC2
phenomenology. Some issues of immediate concern are:

\item{$\circ$} Typical masses of the charged top-pion and its mixing with
technipions. The concern here is is that the decay rates for $t \ra \toppi
b$ or $\tpi b$ may be too large
\ref\tpimix{This concern was first raised by Chris Hill. The problem is
under study by B.~Balaji.}.

\item{$\circ$} Masses of the $\tpi$ and $\trho$. Technipion decays are
mediated by ETC interactions connecting technifermions to quarks and
leptons. Thus, the $\tpi$ are expected to decay to heavy quark and lepton
pairs. The existence of ``leptoquark'' decay modes such as $\tpi \ra b
\ts \tau$ depends on whether ETC operators such as $\ol b_R \gamma^\mu D^1_R
\ol D^l_L \gamma_\mu \ell^h_L$ are allowed. Experiments at the LEP collider
will soon be able to set limits in excess of $75\,\gev$ for charged $\tpi$.
Mixing between gluons and color--octet $\trho$ leads to copious production
of colored $\tpi$; Tevatron collider searches should be able to discover
them with masses up to several hundred~GeV. Production of color--singlet
$\trho \ra \tpi\tpi$, $W_L\tpi$, $Z_L\tpi$, $W_L W_L$, and $W_L Z_L$
should be accessible at the Tevatron for $\tpi$ masses of
$100$--$200\,\gev$~\multi. Another process to be searched for at the
Tevatron is $gg \ra \tpi^0 \ra \ol b b$ or $\ol t t$, if $M_{\pi_T^0} > 2
m_t$. For the longer term, $\trho$ and $\tpi$ masses are needed for LHC and
large $e^+e^-$ collider studies.

\item{$\circ$} Vacuum alignment may produce phases in quark (and
technifermion) mixing matrices that induce detectable CP--violation in the
neutral $K$ and $B$--meson systems, in the neutron electric dipole moment,
and so on. If this happens, it will be important to determine whether
strong CP--violation can be avoided.

\medskip

These brief remarks only scratch the surface of the phenomenological
aspects of the scenario I have presented. I do hope they give a flavor of
the richness of topcolor--assisted technicolor. I do not expect the
specific class of models described here to pass all the tests it faces.
But, in facing them, I expect we will learn how to build more complete and
more successful models.

\bigskip\bigskip

\noindent {\bf Acknowledgements}

I am grateful to Claudio Rebbi for providing a program to solve the
nonlinear equations for the hypercharges. I thank Chris Hill, Estia
Eichten, Sekhar Chivukula, Krishna Kumar and Elizabeth Simmons for helpful
comments. I have benefitted from the hospitality of the Aspen Center for
Physics where this work was begun. This research was supported in part by
the Department of Energy under Grant~No.~DE--FG02--91ER40676.

\vfil\eject

\noindent {\bf Appendix: Anomaly Conditions and Hypercharge Solutions}

\medskip

There are 5~linear and 4~cubic equations for the hypercharges in
Table~1 arising from the requirement that $U(1)_i$ gauge anomalies cancel:
$$\eqalignno{
 U(1)_{1,2} [\sutc]^2:& \quad x_1 - x'_1 + y_1 - y'_1 + z_1 - z'_1 \cr
&\qquad \equiv x'_2 - x_2 + y'_2 - y_2 + z'_2 - z_2
= -\half (\Ntc -2) (\xi - \xi') \cr
 U(1)_{1,2} [\suone]^2:& \quad d = -\Ntc(u_1-v_1) \cr
 U(1)_{1,2} [\sutwo]^2:& \quad b-b' = \half \Ntc(u_1-v_1) \cr
 U(1)_{1,2} [SU(2)]^2:& \quad 2(a+3b) + (c+3d) = -\Ntc [3(u_1 + v_1)
+ x_1 + y_1 + z_1] \cr
&\qquad = \Ntc [3(u_2 + v_2) + x_2 + y_2 + z_2] \cr\cr
 [\uone]^3:& \quad 0= \half \Ntc(\Ntc - 1)(\xi^3 - \xi^{\prime 3}) \cr
& \qquad + 2[2a^3 - a^{\prime 3} + 6(b^3 - b^{\prime 3})] +
             2c^3 - c^{\prime 3} + 6d^3 \cr
& \qquad + 2\Ntc(x^3_1 - x^{\prime 3}_1 + y^3_1 - y^{\prime 3}_1 
+ z^3_1 - z^{\prime 3}_1 ) \cr\cr
 [\utwo]^3:& \quad 0 = -\half \Ntc(\Ntc - 1)(\xi^3 - \xi^{\prime 3}) \cr
& \qquad - 2[2a^3 - a^{\prime 3} + 6(b^3 - b^{\prime 3})] -
             (2c^3 - c^{\prime 3} + 6d^3) \cr
& \qquad + 2\Ntc [x^3_2 - x^{\prime 3}_2 +
y^3_2 - y^{\prime 3}_2 + z^3_2 - z^{\prime 3}_2
 -\textstyle{{9\over{4}}} (u_2+v_2)
 -\textstyle{{3\over{4}}}(x'_2 + y'_2 + z'_2)] \cr
& \qquad + 2[3(a^{\prime 2} - a^2) + 3(a' - \half a)
 + 3(b^2 - b^{\prime 2}) + 5b' - \half b] \cr
& \qquad  +  3(c^{\prime 2} - c^2) + 3(c' - \half c)
 + 3(d^2 - d^{\prime 2}) + 3d' - \half d \cr\cr
 [\uone]^2\utwo:& \quad 0 = -\half \Ntc(\Ntc - 1)(\xi^3 - \xi^{\prime
3})  \cr
&\qquad  - 2[2a^3 - a^{\prime 3} + a^2 - a^{\prime 2}
+ 6(b^3 - b^{\prime 3}) + b^{\prime 2} - b^2] \cr
& \qquad  -(2c^3 - c^{\prime 3} + c^2 - c^{\prime 2}
+ 6d^3  + d^{\prime 2} - d^2) \cr
&\qquad  + 2\Ntc (x^2_1 x_2 - x^{\prime 2}_1 x'_2 +
y^2_1 y_2 - y^{\prime 2}_1 y'_2 +
z^2_1 z_2 - z^{\prime 2}_1 z'_2) \cr\cr
 [\utwo]^2\uone:& \quad 0 = \half \Ntc(\Ntc - 1)(\xi^3 - \xi^{\prime 3})
\cr
& \qquad + 2[2a^3 - a^{\prime 3} + 2(a^2 - a^{\prime 2}) 
 + 6(b^3 - b^{\prime 3}) + 2(b^{\prime 2} - b^2)] \cr
& \qquad +  2c^3 -  c^{\prime 3} + 2(c^2 - c^{\prime 2})
 +  6d^3 + 2(d^{\prime 2} - d^2) \cr
& \qquad + 2\Ntc [x^2_2 x_1 - x^{\prime 2}_2 x'_1 +
y^2_2 y_1 - y^{\prime 2}_2 y'_1 + z^2_2 z_1 - z^{\prime 2}_2 z'_1 \cr
& \qquad -\textstyle{{3\over{4}}} (u_1+v_1)
 -\textstyle{{1\over{4}}}(x'_1 + y'_1 + z'_1)] \cr 
& \qquad + 2(\half a - a' + \sixth b -  \textstyle{{5\over{3}}} b')
 +  \half c - c' + \sixth d - d'
\ts\ts . &(A.1) \cr}
$$
These 4~cubic equations are not independent because the $[\uy]^3 =
[\uone + \utwo]^3$ anomaly cancellation is guaranteed by the $\uy
[SU(2)]^2$ condition.
A convenient set of 3~independent cubic equations consists of
$[\uone]^3$ plus $[\uone]^2 \uy$ and $[\uone]^3 + [\utwo]^3 - 3 [\uone]^2
\uy$:
$$\eqalignno{
 [\uone]^2 \uy:& \quad 0 = 2(a^{\prime 2} - a^2 + b^2 - b^{\prime 2})
+ c^{\prime 2} - c^2  + d^2 - d^{\prime 2} \cr
& \qquad +  2\Ntc [(x_1 + x_2) (x^2_1 - x^{\prime 2}_1) 
+ (y_1 + y_2) (y^2_1 - y^{\prime 2}_1) 
+ (z_1 + z_2) (z^2_1 - z^{\prime 2}_1)] \cr\cr
[\uone]^3 +  [\utwo]^3 &  - 3 [\uone]^2 \uy:\cr
& \quad 0 = (u_1-v_1) \{2\Ntc^2 [4(y_1 + y_2)^2 - (x_1 + x_2)^2
+ \textstyle{{3\over{4}}}] - (5\Ntc + 2)\} \ts\ts. &(A.2) \cr}
$$
In the last equation, I used results from Eq.~\charges.

The 18~linear and 3~nonlinear equations satisfied by the 26~hypercharges do
not determine them uniquely. I sought numerical solutions to them that have
$u = \half(u_1 - v_1) \neq 0$ as follows: First, I set $\xi' = - \xi$ and
$c = a$. Then I chose values for $x_1$, $y_1$ and $y_1 + y_2$, and solved
for $u$, $a$ and $x_1 + x_2$. To obtain $u_1 - v_1 = \CO(1)$, I
input $x_1$, $y_1 = \CO(\Ntc u)$. For $\Ntc = 4$ and $y_1 + y_2 = 0$ (which
implies $x_1 + x_2 = \pm \fourth$) and $x_1 = y_1 = 10$, I obtained
$$\eqalignno{
& u = 1.075 \ts, \quad a = 1.040 \qquad ({\rm for} \ts\ts
x_1 + x_2 = -\fourth) \cr
& u = 1.197 \ts, \quad a = 12.054 \qquad ({\rm for} \ts\ts
x_1 + x_2 = \fourth) \ts\ts. &(A.3) \cr}
$$
As is apparent from Eqs.~(A.2), these solutions scale linearly with the
input values of $x_1$ and $y_1$. Values of $a$ as large as~12 are
doubtless ruled out.

\listrefs

\centerline{\vbox{\offinterlineskip
\hrule\hrule\hrule
\halign{&\vrule#&
  \strut\quad#\hfil\quad\cr
height4pt&\omit&&\omit&&\omit&&\omit&&\omit&\cr\cr
&\hfill Particle \hfill&&\hfill$SU(3)_1$ \hfill&&\hfill
 $SU(3)_2$\hfill&&\hfill$Y_1$\hfill&&\hfill $Y_2$\hfill &\cr\cr
height4pt&\omit&&\omit&&\omit&&\omit&&\omit&\cr\cr
\noalign{\hrule\hrule\hrule}
height4pt&\omit&&\omit&&\omit&&\omit&&\omit&\cr\cr
&$\ell_L^l$&&\hfill$1$\hfill&&\hfill$1$\hfill
&&\hfill$a$\hfill&&\hfill$-\half-a$\hfill&\cr\cr
\noalign{}
height4pt&\omit&&\omit&&\omit&&\omit&&\omit&\cr\cr
&$e_R$, $\mu_R$&&\hfill$1$\hfill&&\hfill$1$\hfill
&&\hfill$a'$\hfill&&\hfill$-1-a'$\hfill&\cr\cr
\noalign{\hrule}
height4pt&\omit&&\omit&&\omit&&\omit&&\omit&\cr\cr
&$q_L^l$&&\hfill$1$\hfill&&\hfill$3$\hfill
&&\hfill$b$\hfill&&\hfill$\sixth-b$\hfill&\cr\cr
\noalign{}
height4pt&\omit&&\omit&&\omit&&\omit&&\omit&\cr\cr
&$u_R$, $c_R$&&\hfill$1$\hfill&&\hfill$3$\hfill
&&\hfill$b'$\hfill&&\hfill$\twothirds-b'$\hfill&\cr\cr
\noalign{}
height4pt&\omit&&\omit&&\omit&&\omit&&\omit&\cr\cr
&$d_R$, $s_R$&&\hfill$1$\hfill&&\hfill$3$\hfill
&&\hfill$b'$\hfill&&\hfill$-\third - b'$\hfill&\cr\cr
\noalign{\hrule}
height4pt&\omit&&\omit&&\omit&&\omit&&\omit&\cr\cr
&$\ell_L^h$&&\hfill$1$\hfill&&\hfill$1$\hfill
&&\hfill$c$\hfill&&\hfill$-\half-c$\hfill&\cr\cr
\noalign{}
height4pt&\omit&&\omit&&\omit&&\omit&&\omit&\cr\cr
&$\tau_R$&&\hfill$1$\hfill&&\hfill$1$\hfill
&&\hfill$c'$\hfill&&\hfill$-1-c'$\hfill&\cr\cr
\noalign{\hrule}
height4pt&\omit&&\omit&&\omit&&\omit&&\omit&\cr\cr
&$q_L^h$&&\hfill$3$\hfill&&\hfill$1$\hfill
&&\hfill$d$\hfill&&\hfill$\sixth-d$\hfill&\cr\cr
\noalign{}
height4pt&\omit&&\omit&&\omit&&\omit&&\omit&\cr\cr
&$t_R$&&\hfill$3$\hfill&&\hfill$1$\hfill
&&\hfill$d'$\hfill&&\hfill$\twothirds-d'$\hfill&\cr\cr
\noalign{}
height4pt&\omit&&\omit&&\omit&&\omit&&\omit&\cr\cr
&$b_R$&&\hfill$3$\hfill&&\hfill$1$\hfill
&&\hfill$-d'$\hfill&&\hfill$-\third + d'$\hfill&\cr\cr
\noalign{\hrule\hrule}
height4pt&\omit&&\omit&&\omit&&\omit&&\omit&\cr\cr
&$T_L^1$&&\hfill$3$\hfill&&\hfill$1$\hfill
&&\hfill$u_1$\hfill&&\hfill$u_2$\hfill&\cr\cr
\noalign{}
height4pt&\omit&&\omit&&\omit&&\omit&&\omit&\cr\cr
&$U_R^1$&&\hfill$3$\hfill&&\hfill$1$\hfill
&&\hfill$v_1$\hfill&&\hfill$v_2+\half$\hfill&\cr\cr
\noalign{}
height4pt&\omit&&\omit&&\omit&&\omit&&\omit&\cr\cr
&$D_R^1$&&\hfill$3$\hfill&&\hfill$1$\hfill
&&\hfill$v_1$\hfill&&\hfill$v_2-\half$\hfill&\cr\cr
\noalign{\hrule}
height4pt&\omit&&\omit&&\omit&&\omit&&\omit&\cr\cr
&$T_L^2$&&\hfill$1$\hfill&&\hfill$3$\hfill
&&\hfill$v_1$\hfill&&\hfill$v_2$\hfill&\cr\cr
\noalign{}
height4pt&\omit&&\omit&&\omit&&\omit&&\omit&\cr\cr
&$U_R^2$&&\hfill$1$\hfill&&\hfill$3$\hfill
&&\hfill$u_1$\hfill&&\hfill$u_2+\half$\hfill&\cr\cr
\noalign{}
height4pt&\omit&&\omit&&\omit&&\omit&&\omit&\cr\cr
&$D_R^2$&&\hfill$1$\hfill&&\hfill$3$\hfill
&&\hfill$u_1$\hfill&&\hfill$u_2-\half$\hfill&\cr\cr
\noalign{\hrule\hrule}
height4pt&\omit&&\omit&&\omit&&\omit&&\omit&\cr\cr
&$T_L^l$&&\hfill$1$\hfill&&\hfill$1$\hfill
&&\hfill$x_1$\hfill&&\hfill$x_2$\hfill&\cr\cr
\noalign{}
height4pt&\omit&&\omit&&\omit&&\omit&&\omit&\cr\cr
&$U_R^l$&&\hfill$1$\hfill&&\hfill$1$\hfill
&&\hfill$x'_1$\hfill&&\hfill$x'_2+\half$\hfill&\cr\cr
\noalign{}
height4pt&\omit&&\omit&&\omit&&\omit&&\omit&\cr\cr
&$D_R^l$&&\hfill$1$\hfill&&\hfill$1$\hfill
&&\hfill$x'_1$\hfill&&\hfill$x'_2-\half$\hfill&\cr\cr
\noalign{\hrule}
height4pt&\omit&&\omit&&\omit&&\omit&&\omit&\cr\cr
&$T_L^t$&&\hfill$1$\hfill&&\hfill$1$\hfill
&&\hfill$y_1$\hfill&&\hfill$y_2$\hfill&\cr\cr
\noalign{}
height4pt&\omit&&\omit&&\omit&&\omit&&\omit&\cr\cr
&$U_R^t$&&\hfill$1$\hfill&&\hfill$1$\hfill
&&\hfill$y'_1$\hfill&&\hfill$y'_2+\half$\hfill&\cr\cr
\noalign{}
height4pt&\omit&&\omit&&\omit&&\omit&&\omit&\cr\cr
&$D_R^t$&&\hfill$1$\hfill&&\hfill$1$\hfill
&&\hfill$y'_1$\hfill&&\hfill$y'_2-\half$\hfill&\cr\cr
\noalign{\hrule}
height4pt&\omit&&\omit&&\omit&&\omit&&\omit&\cr\cr
&$T_L^b$&&\hfill$1$\hfill&&\hfill$1$\hfill
&&\hfill$z_1$\hfill&&\hfill$z_2$\hfill&\cr\cr
\noalign{}
height4pt&\omit&&\omit&&\omit&&\omit&&\omit&\cr\cr
&$U_R^b$&&\hfill$1$\hfill&&\hfill$1$\hfill
&&\hfill$z'_1$\hfill&&\hfill$z'_2+\half$\hfill&\cr\cr
\noalign{}
height4pt&\omit&&\omit&&\omit&&\omit&&\omit&\cr\cr
&$D_R^b$&&\hfill$1$\hfill&&\hfill$1$\hfill
&&\hfill$z'_1$\hfill&&\hfill$z'_2-\half$\hfill&\cr\cr
\noalign{\hrule\hrule}
height4pt&\omit&&\omit&&\omit&&\omit&&\omit&\cr\cr
&$\psi_L$&&\hfill$1$\hfill&&\hfill$1$\hfill
&&\hfill$\xi$\hfill&&\hfill$-\xi$\hfill&\cr\cr
\noalign{}
height4pt&\omit&&\omit&&\omit&&\omit&&\omit&\cr\cr
&$\psi_R$&&\hfill$1$\hfill&&\hfill$1$\hfill
&&\hfill$\xi'$\hfill&&\hfill$-\xi'$\hfill&\cr\cr
height4pt&\omit&&\omit&&\omit&&\omit&&\omit&\cr\cr}
\hrule\hrule\hrule}}

\centerline{TABLE 1: Lepton, quark and technifermion colors and
hypercharges.}

\vfil\eject

\bye